\batchmode
\makeatletter
\def\input@path{{"/home/jacob/Documents/Work/My Papers/2026-Against Many Worlds/"}}
\makeatother
\documentclass[11pt,twoside,english]{article}
\usepackage[T1]{fontenc}
\usepackage[latin9]{inputenc}
\usepackage{color}
\definecolor{lyxboxbgcolor}{rgb}{0.980469, 0.941406, 0.902344}
\usepackage{babel}
\usepackage{amsmath}
\usepackage{amssymb}
\usepackage{geometry}
\usepackage{setspace}
\usepackage[pdftex,pdfusetitle,
 bookmarks=true,bookmarksnumbered=true,bookmarksopen=false,
 breaklinks=false,pdfborder={0 0 0},pdfborderstyle={},backref=false,colorlinks=false]
 {hyperref}

\makeatletter

\let\originalleft\left
\let\originalright\right
\renewcommand{\left}{\mathopen{}\mathclose\bgroup\originalleft}
\renewcommand{\right}{\aftergroup\egroup\originalright}

\def\smalloverbrace#1{\mathop{\vbox{\m@th\ialign{##\crcr%
      \noalign{\kern3\p@}%
      \tiny\downbracefill\crcr\noalign{\kern3\p@\nointerlineskip}%
      $\hfil\displaystyle{#1}\hfil$\crcr}}}\limits}

\def\smallunderbrace#1{\mathop{\vtop{\m@th\ialign{##\crcr
   $\hfil\displaystyle{#1}\hfil$\crcr
   \noalign{\kern3\p@\nointerlineskip}%
   \tiny\upbracefill\crcr\noalign{\kern3\p@}}}}\limits}

\DeclareMathAlphabet{\mymathbb}{U}{bbold}{m}{n}

\makeatother

\begin{document}
\title{Against Many Worlds}
\author{Emily Adlam\thanks{Department of Philosophy and Institute for Quantum Studies, Chapman University, 1 University Drive, Orange, CA 92866; adlam@chapman.edu; ORCID: 0000-0002-5998-7685},
Jacob A. Barandes\thanks{Departments of Philosophy and Physics, Harvard University, Cambridge, MA 02138; jacob\_barandes@harvard.edu; ORCID: 0000-0002-3740-4418}
}
\date{}

\maketitle

\begin{abstract}
Any viable interpretation of quantum theory needs to account for the
Born rule, from which the theory gets its probabilistic empirical
predictions. In this paper, we give an overview of possible approaches
to this problem in the context of the Many Worlds interpretation.
We argue that, for structural reasons, none of them can possibly succeed.
More precisely, we argue that the Many Worlds interpretation must
obtain the Born rule by proceeding either axiomatically, deductively,
or inductively, and that all three of these approaches run into general,
fundamental obstructions.
\end{abstract}

\begin{center}
\global\long\def\quote#1{``#1"}%
\global\long\def\apostrophe{\textrm{'}}%
\global\long\def\slot{\phantom{x}}%
\global\long\def\eval#1{\left.#1\right\vert }%
\global\long\def\keyeq#1{\boxed{#1}}%
\global\long\def\importanteq#1{\boxed{\boxed{#1}}}%
\global\long\def\given{\vert}%
\global\long\def\mapping#1#2#3{#1:#2\to#3}%
\global\long\def\composition{\circ}%
\global\long\def\set#1{\left\{  #1\right\}  }%
\global\long\def\setindexed#1#2{\left\{  #1\right\}  _{#2}}%

\global\long\def\setbuild#1#2{\left\{  \left.\!#1\,\right|\,#2\right\}  }%
\global\long\def\complem{\mathrm{c}}%

\global\long\def\union{\cup}%
\global\long\def\intersection{\cap}%
\global\long\def\cartesianprod{\times}%
\global\long\def\disjointunion{\sqcup}%

\global\long\def\isomorphic{\cong}%

\global\long\def\setsize#1{\left|#1\right|}%
\global\long\def\defeq{\equiv}%
\global\long\def\conj{\ast}%
\global\long\def\overconj#1{\overline{#1}}%
\global\long\def\re{\mathrm{Re\,}}%
\global\long\def\im{\mathrm{Im\,}}%

\global\long\def\transp{\mathrm{T}}%
\global\long\def\tr{\mathrm{tr}}%
\global\long\def\adj{\dagger}%
\global\long\def\diag#1{\mathrm{diag}\left(#1\right)}%
\global\long\def\dotprod{\cdot}%
\global\long\def\crossprod{\times}%
\global\long\def\Probability#1{\mathrm{Prob}\left(#1\right)}%
\global\long\def\Amplitude#1{\mathrm{Amp}\left(#1\right)}%
\global\long\def\cov{\mathrm{cov}}%
\global\long\def\corr{\mathrm{corr}}%

\global\long\def\absval#1{\left\vert #1\right\vert }%
\global\long\def\expectval#1{\left\langle #1\right\rangle }%
\global\long\def\op#1{\hat{#1}}%

\global\long\def\bra#1{\left\langle #1\right|}%
\global\long\def\ket#1{\left|#1\right\rangle }%
\global\long\def\braket#1#2{\left\langle \left.\!#1\right|#2\right\rangle }%

\global\long\def\parens#1{(#1)}%
\global\long\def\bigparens#1{\big(#1\big)}%
\global\long\def\Bigparens#1{\Big(#1\Big)}%
\global\long\def\biggparens#1{\bigg(#1\bigg)}%
\global\long\def\Biggparens#1{\Bigg(#1\Bigg)}%
\global\long\def\bracks#1{[#1]}%
\global\long\def\bigbracks#1{\big[#1\big]}%
\global\long\def\Bigbracks#1{\Big[#1\Big]}%
\global\long\def\biggbracks#1{\bigg[#1\bigg]}%
\global\long\def\Biggbracks#1{\Bigg[#1\Bigg]}%
\global\long\def\curlies#1{\{#1\}}%
\global\long\def\bigcurlies#1{\big\{#1\big\}}%
\global\long\def\Bigcurlies#1{\Big\{#1\Big\}}%
\global\long\def\biggcurlies#1{\bigg\{#1\bigg\}}%
\global\long\def\Biggcurlies#1{\Bigg\{#1\Bigg\}}%
\global\long\def\verts#1{\vert#1\vert}%
\global\long\def\bigverts#1{\big\vert#1\big\vert}%
\global\long\def\Bigverts#1{\Big\vert#1\Big\vert}%
\global\long\def\biggverts#1{\bigg\vert#1\bigg\vert}%
\global\long\def\Biggverts#1{\Bigg\vert#1\Bigg\vert}%
\global\long\def\Verts#1{\Vert#1\Vert}%
\global\long\def\bigVerts#1{\big\Vert#1\big\Vert}%
\global\long\def\BigVerts#1{\Big\Vert#1\Big\Vert}%
\global\long\def\biggVerts#1{\bigg\Vert#1\bigg\Vert}%
\global\long\def\BiggVerts#1{\Bigg\Vert#1\Bigg\Vert}%
\global\long\def\ket#1{\vert#1\rangle}%
\global\long\def\bigket#1{\big\vert#1\big\rangle}%
\global\long\def\Bigket#1{\Big\vert#1\Big\rangle}%
\global\long\def\biggket#1{\bigg\vert#1\bigg\rangle}%
\global\long\def\Biggket#1{\Bigg\vert#1\Bigg\rangle}%
\global\long\def\bra#1{\langle#1\vert}%
\global\long\def\bigbra#1{\big\langle#1\big\vert}%
\global\long\def\Bigbra#1{\Big\langle#1\Big\vert}%
\global\long\def\biggbra#1{\bigg\langle#1\bigg\vert}%
\global\long\def\Biggbra#1{\Bigg\langle#1\Bigg\vert}%
\global\long\def\braket#1#2{\langle#1\vert#2\rangle}%
\global\long\def\bigbraket#1#2{\big\langle#1\big\vert#2\big\rangle}%
\global\long\def\Bigbraket#1#2{\Big\langle#1\Big\vert#2\Big\rangle}%
\global\long\def\biggbraket#1#2{\bigg\langle#1\bigg\vert#2\bigg\rangle}%
\global\long\def\Biggbraket#1#2{\Bigg\langle#1\Bigg\vert#2\Bigg\rangle}%
\global\long\def\angs#1{\langle#1\rangle}%
\global\long\def\bigangs#1{\big\langle#1\big\rangle}%
\global\long\def\Bigangs#1{\Big\langle#1\Big\rangle}%
\global\long\def\biggangs#1{\bigg\langle#1\bigg\rangle}%
\global\long\def\Biggangs#1{\Bigg\langle#1\Bigg\rangle}%

\global\long\def\vec#1{\mathbf{#1}}%
\global\long\def\vecgreek#1{\boldsymbol{#1}}%
\global\long\def\idmatrix{\mymathbb{1}}%
\global\long\def\projector{P}%
\global\long\def\permutationmatrix{\Sigma}%
\global\long\def\densitymatrix{\rho}%
\global\long\def\krausmatrix{K}%
\global\long\def\stochasticmatrix{\Gamma}%
\global\long\def\lindbladmatrix{L}%
\global\long\def\dynop{\Theta}%
\global\long\def\timeevop{U}%
\global\long\def\hadamardprod{\odot}%
\global\long\def\tensorprod{\otimes}%

\global\long\def\inprod#1#2{\left\langle #1,#2\right\rangle }%
\global\long\def\normket#1{\left\Vert #1\right\Vert }%
\global\long\def\hilbspace{\mathcal{H}}%
\global\long\def\samplespace{\Omega}%
\global\long\def\configspace{\mathcal{C}}%
\global\long\def\phasespace{\mathcal{P}}%
\global\long\def\spectrum{\sigma}%
\global\long\def\restrict#1#2{\left.#1\right\vert _{#2}}%
\global\long\def\from{\leftarrow}%
\global\long\def\statemap{\omega}%
\global\long\def\degangle#1{#1^{\circ}}%
\global\long\def\trivialvector{\tilde{v}}%
\global\long\def\eqsbrace#1{\left.#1\qquad\right\}  }%
\global\long\def\operator#1{\operatorname{#1}}%
\par\end{center}

\section{Introduction\label{sec:Introduction}}

The Many Worlds interpretation of quantum theory was originally proposed
by Hugh Everett in the mid-1950s (Everett 1956, 1957)\nocite{Everett:1956ttotuwf,Everett:1957rsfqm}
and now appears in several different formulations. Our primary goal
in this paper is to argue that no formulation of the Many Worlds interpretation
is capable of meeting the challenge of accounting for a pivotal ingredient
of quantum theory called the Born rule. If this is so, the Many Worlds
interpretation cannot be a viable candidate for an empirically adequate
interpretation of quantum theory.

Similar criticisms have been raised many times in the research literature
(e.g., Kent 1990, Putnam 2005, Albert 2010, Adlam 2014)\nocite{Kent:1990amwi,Putnam:2005aplaqma,Albert:2010pitep,Adlam:2014tpocitei}.
Our goal here is to take a wider view, identifying basic structural
problems with the Many Worlds interpretation that underlie many of
those previous criticisms. In summary, our strategy will be to argue
that any admissible attempt to justify or explain the Born rule must
take one of three possible approaches \textendash{} axiomatic, deductive,
or inductive \textendash{} and that all three approaches, as well
as mixtures of them, run into insurmountable obstructions if one works
within the Many Worlds interpretation.

In outline, we will proceed as follows: In Section~\ref{sec:The-Born-Rule},
we will introduce the Born rule. In Section~\ref{sec:The-Many-Worlds-Interpretation},
we will introduce the Many Worlds interpretation. In Section~\ref{sec:Summary-of-the-Case},
we will present a concise summary of our overall dialectic. In Sections~\ref{sec:The-Axiomatic-Approach},
\ref{sec:The-Deductive-Approach}, and \ref{sec:The-Inductive-Approach},
respectively, we will present detailed examinations of the axiomatic,
deductive, and inductive approaches to justifying the Born rule on
the Many Worlds interpretation. In Section~\ref{sec:A-Fourth-Way?},
we will consider whether there could be any alternative approaches.
We will conclude in Section~\ref{sec:Conclusion} with a summary
and a discussion of future research directions.

\section{The Born Rule\label{sec:The-Born-Rule}}

The Born rule (Born 1926)\nocite{Born:1926zqds} is an essential feature
of quantum theory because it is the main connection between the theoretical
machinery of the theory and the experimental results that are supposed
to serve as evidence for it. In short, the Born rule says that if
we take two abstract ingredients \textendash{} the mathematical representation
$\rho$ of a system\textquoteright s quantum state and a projection
operator $P$ representing a particular measurement outcome \textendash{}
and plug them into a very specific formula, $\operatorname{Trace}(P\rho)$,
then this formula gives the probability for $P$ to occur if an observer
performs a measurement on this system for which $P$ is a possible
outcome. In the simplest case, where $\rho$ can be expressed in terms
of a wave function $\vert\Psi\rangle$ and $P$ is an operator projecting
onto a single vector or \textquoteleft pure state,\textquoteright{}
the Born rule amounts to taking that wave function $\vert\Psi\rangle$,
decomposing it into a superposition of vector-components $\vert v\rangle,\vert w\rangle,\dots$
that represent different possible measurement outcomes, 
\[
\vert\Psi\rangle=\alpha\vert v\rangle+\beta\vert w\rangle+\cdots,
\]
 where $\alpha,\beta,\dots$ are complex-valued numerical coefficients,
and then setting the probability of each possible measurement outcome
to be the \textquoteleft mod-square\textquoteright{} ${\vert\alpha\vert}^{2},{\vert\beta\vert}^{2},\dots$
of the appropriate complex-valued coefficient. For example, the probability
of obtaining the outcome corresponding to $\vert v\rangle$ for the
wave function $\vert\Psi\rangle$ given above is simply ${\vert\alpha\vert}^{2}$.

The Born rule is a reliable, robust, quantitatively precise, predictive
generalization of patterns of observation about highly generic kinds
of microscopic systems. As inputs, the Born rule takes information
about the past preparation of the system\textquoteright s state together
with the system\textquoteright s Hamiltonian \textendash{} a mathematical
operator that describes the state\textquoteright s time evolution.
Then, as an output, the Born rule fixes probability distributions
for future measurement outcomes. Those probability distributions and
their downstream consequences have been verified empirically countless
times over the past century, in some cases to 13 decimal places (Fan
et al. 2023)\nocite{FanMyersSukraGabrielse:2023motemm}. 

We leave it to the reader to decide whether the Born rule should qualify
as the sort of theoretical ingredient that could be regarded as a
law of nature \textendash{} and, for that matter, whether there are
any known laws of nature in all of science that are in better accord
with empirical data. The status of the Born rule as a candidate law
turns on a number of unresolved debates in the metaphysics of laws
(Carroll 2025)\nocite{Carroll:2025lon}. We will return to this
question in the concluding section of this paper. 

What is not in question is the Born rule\textquoteright s distinctly
lawlike set of properties, as listed above. As such, we will generally
refer to the Born rule as a \textquoteleft lawlike ingredient\textquoteright{}
of quantum theory.

Broadly speaking, all of the empirical predictions of quantum theory
descend either from the Born rule, or from numbers (called eigenvalues)
extracted from the abstract operators that represent observables in
the theory \textendash{} such as the numbers characterizing a quantum
system\textquoteright s energy levels. In particular, without the
Born rule, much of the empirical content of quantum theory would be
missing. Indeed, as a twist on Whitehead, one could say that the entire
content of the Particle Data Group\textquoteright s reference booklet
consists of a series of footnotes to the Born rule.\footnote{See \href{https://pdg.lbl.gov/}{https://pdg.lbl.gov/}.}
Any interpretation, formulation, or alternative theory of quantum
mechanics that purports to be empirically adequate is therefore required
to account for the Born rule in some way.

\section{The Many Worlds Interpretation\label{sec:The-Many-Worlds-Interpretation}}

The Many Worlds interpretation posits that the physical state of the
entire universe is represented by a \textquoteleft universal wave
function\textquoteright{} evolving according to a specific, deterministic
rule, called unitary time evolution. Under commonly assumed continuity
conditions, unitary time evolution can be reduced to a first-order
differential equation, called the Schrödinger equation, that involves
a mathematical operator called a Hamiltonian.

A nomologically contingent claim of the Many Worlds interpretation
is that for our actual universe, the universal wave function, as a
formal vector in a very high-dimensional vector space called a Hilbert
space, naturally decomposes over time into more and more vector-components
that barely interfere with each other. When these vector-components
stop appreciably interfering with each other, through a process called
decoherence, they are called \textquoteleft macro-world branches,\textquoteright{}
or just \textquoteleft branches\textquoteright{} for short.

Putting aside some separately contentious issues regarding representation
and emergence, a central claim of the Many Worlds interpretation is
that within each such macro-world branch, one finds a representation
of what might be regarded as a complete classical reality. In particular,
the claim is that at least some of these macro-world branches \emph{qua}
classical realities are filled with galaxies and planets and observers.

If the Many Worlds interpretation is to succeed in accounting for
the Born rule, then measurement probabilities must somehow be associated
with the complex-valued coefficients sitting in front of the vector-components
representing macro-world branches. The challenge of establishing this
association for the Many Worlds interpretation is called the \textquoteleft problem
of probability.\textquoteright{}

It is common for debates over the problem of probability to devolve
into narrow, highly technical discussions about such issues as diachronic
consistency, Dutch-book coherence, unitary operations, coarse-graining,
Hilbert-space measures, and the like (see, e.g., Wallace 2012, Carroll
2019)\nocite{Wallace:2012temqtattei,Carroll:2019sdhqwateos}. These
sorts of approaches have yielded valuable insights, but sometimes
risk missing the forest for the trees. 

We will therefore try to avoid getting lost in the weeds by keeping
the discussion in this paper at a high level, focusing on general
arguments that identify fundamental obstacles to getting the Many
Worlds interpretation to work. To those ends, we will argue that the
Many Worlds interpretation fails right at the starting gate by running
into more basic, first-order obstructions at a level akin to, say,
Hume\textquoteright s argument for the impossibility of deriving an
\textquoteleft ought\textquoteright{} from an \textquoteleft is.\textquoteright \footnote{Fittingly, in a 1999 paper introducing decision-theoretic arguments
that would ultimately become a standard feature of the \textquoteleft Oxford
school\textquoteright{} of the Many Worlds interpretation, Deutsch
argued that he could overcome just this sort of first-order obstruction,
boldly claiming: \textquotedblleft Thus we see that quantum theory
permits what philosophy would hitherto have regarded as a formal impossibility,
akin to \textquoteleft deriving an ought from an is\textquoteright ,
namely deriving a probability statement from a factual statement.
This could be called deriving a \textquoteleft tends to\textquoteright{}
from a \textquoteleft does\textquoteright .\textquotedblright{} (Deutsch
1999)\nocite{Deutsch:1999qtopad}}

Solving the problem of probability is not a trivial matter. For one
thing, it is not sufficient merely to show that the modulus-squares
of the coefficients, known as branch-weights, are non-negative numbers
that sum up to one. Nor is it sufficient to argue that some suitable
construct based on Hilbert spaces satisfies the mathematical axioms
of a Kolmogorov probability space. After all, the world contains many
things that satisfy these mathematical axioms but are clearly not
probabilities, such as mass density distributions.

Moreover, it is likewise not adequate to count up macroworld branches
and appeal to the common-sense notion that all branches should have
the same probability. One obstruction is that \textquoteleft counting
up\textquoteright{} branches is not a mathematically well-defined
notion, as advocates of the Many Worlds interpretation readily acknowledge
(Wallace 2012)\nocite{Wallace:2012temqtattei}. Another obstruction
is that the most natural way of counting does not in any case yield
the correct quantitative probabilities required by the Born rule (Saunders
2021)\nocite{Saunders:2021bciteioqm}.

A more fundamental obstruction is that merely having multiple things
is not enough, by itself, to imply a conception of probability. To
see why, consider, for example, a jar containing 43 red marbles among
a total of 100 marbles. On the one hand, it would be entirely reasonable
to say that the fraction of red marbles was 43\%. On the other hand,
it would not yet be reasonable to call this fraction a \emph{probability},
because, at this point in the thought experiment, the probability
would not be a probability \emph{of} anything. It would not make sense
to say that there was a 43\% probability \emph{simpliciter}.

Of course, if we were to stipulate that an agent were picking a marble
\textquoteleft at random\textquoteright{} from the jar, under some
suitable notion of \textquoteleft at random\textquoteright{} and with
the further assumption that the marbles were \textquoteleft well-mixed,\textquoteright{}
then we would be in a reasonable position to assert that there was
a probability of 43\% \emph{of the agent picking a red marble}. But
absent such a picking process, or the introduction of some other selection
process to serve a similar functional role in the argument, there
would be no probability, because, again, there would simply be nothing
for the probability to be a probability \emph{of}.

It is worth dwelling for a moment more on what premises are needed
for a \textquoteleft picking process\textquoteright{} to legitimate
the move from 43\% \emph{qua} fraction to 43\% \emph{qua} probability.
Assigning a picking probability of 43\% amounts to adopting a \textquoteleft principle
of indifference\textquoteright{} (Eva 2019)\nocite{Eva:2019poi} assuming
that all the marbles are \textquoteleft equally likely\textquoteright{}
to be picked. And the philosophical literature on indifference principles
makes clear that they are justified and successful only when they
reflect relevant features of the process by which an outcome is selected,
such as the symmetries of the picking process (van Fraassen 1989,
Shackel 2007)\nocite{VanFraassen:1989las,Shackel:2007bspatpoi}.

In particular, assigning a probability of 43\% makes sense if the
agent responsible for picking a marble is doing so by some process
that is blind to color, meaning that there is a symmetry with respect
to color \textendash{} hence our invocation of terms like \textquoteleft at
random\textquoteright{} and \textquoteleft well-mixed.\textquoteright{}
Obviously if the agent is picking marbles in some way that is biased
toward red marbles, then it would not make sense to assign a probability
of 43\%.

This demonstrates that arriving at a meaningful probability assignment
requires specifying or making assumptions about the details of a selection
process, such as the symmetries of that selection process. Thus we
cannot arrive at \emph{a priori} probabilities based on a description
of a physical scenario without any kind of selection process at all.
This means we cannot sensibly arrive at probabilities at all if there
\emph{is} no selection process, and that is exactly the situation
for the Many Worlds interpretation, since it is essential to the Many
Worlds picture that all of the branches are always produced and there
is no structure in the world picking out a privileged branch.

One might wonder whether replacing marbles with entire classical realities
\textendash{} each containing observers \textendash{} might be a different
sort of modification that licenses talk of probabilities. \emph{Prima
facie}, it might seem reasonable to think that such observers could
be uncertain about their location among these classical realities,
and that \textquoteleft self-locating\textquoteright{} uncertainties
of this kind could be given a probabilistic meaning in terms of subjective
credences. In the discussion to follow, we will argue that this move,
which lies at the core of the Many Worlds interpretation, cannot successfully
yield the Born rule.

\section{Summary of the Case\label{sec:Summary-of-the-Case}}

The dialectic ahead will take the following schematic form.

First, we will argue that any lawlike ingredient of a physical theory
needs to be justified in one of three ways.
\begin{enumerate}
\item The \textquoteleft axiomatic\textquoteright{} approach, which proceeds
by postulating the lawlike ingredient outright as one of the theory\textquoteright s
\textquoteleft physical\textquoteright{} axioms. Examples of such
axioms include Newton\textquoteright s second law, $F=ma$, or the
Maxwell equations of classical electromagnetism, or the Many Worlds
interpretation\textquoteright s law of unitary time evolution for
the universal wave function. Axioms of this kind are purely descriptive
in nature and do not include, for instance, normative principles of
scientific inquiry or reasoning upheld by \textquoteleft rational\textquoteright{}
observers.
\item The \textquoteleft deductive\textquoteright{} approach, which consists
of arriving at a lawlike ingredient by an argument from the theory\textquoteright s
physical axioms, perhaps together with other \textquoteleft reasonable\textquoteright{}
assumptions or normative principles that are consistent with the theory\textquoteright s
physical axioms. Examples would include conservation laws derived
from a theory\textquoteright s physical axioms by way of Noether\textquoteright s
theorem. Another example would be Kepler\textquoteright s laws of
planetary motion, derived from Newton\textquoteright s mechanics and
his theory of universal gravitation.
\item The \textquoteleft inductive\textquoteright{} approach, meaning an
application of inductive reasoning without appeal to any intermediate
theoretical structure, such as by asserting that objects with mass
will fall in a \textquoteleft specific manner\textquoteright{} in
the future based on observations that they have reliably fallen in
a \textquoteleft similar\textquoteright{} manner in the past. We readily
acknowledge the vagueness of our specification of the inductive approach,
including the terms \textquoteleft specific manner\textquoteright{}
and \textquoteleft similar.\textquoteright{} This vagueness is a deliberate
attempt to be as generous and accommodating as possible to advocates
of the Many Worlds interpretation.
\end{enumerate}
We will consider each of these three general approaches in turn, and
argue that the structure of the Many Worlds interpretation blocks
each of these approaches from succeeding as a means of deriving the
Born rule. We will preview those three arguments here, and then lay
them out in more detail in the following sections.

\subsection{The Case Against the Axiomatic Approach\label{subsec:The-Case-Against-the-Axiomatic-Approach}}

The basic problem with the axiomatic approach is that most axiomatic
formulations of the Many Worlds interpretation (Everett 1956, DeWitt
1970, Deutsch 1999, Wallace 2012, Saunders 2021, Carroll 2019)\nocite{Everett:1956ttotuwf,DeWitt:1970qmr,Deutsch:1999qtopad,Wallace:2012temqtattei,Saunders:2021bciteioqm,Carroll:2019sdhqwateos}
do not include physical axioms that introduce any \emph{fundamental}
physical ingredients or referents to which the Born rule could attach.
In particular, on modern understandings of the Many Worlds interpretation,
the Born rule is supposed to refer to macro-world branches that appear
\emph{secondarily} to the axioms, as emergent or derived physical
structures arising from decoherence in a contingent way. Observers
and measuring devices are similarly secondary (or perhaps tertiary)
physical structures.

One cannot assign \emph{axiomatic} properties to these derived structures,
any more than one could derive tables and chairs from a theory of
chemistry and then axiomatically stipulate that these tables and chairs
must turn out to be yellow \textendash{} the properties of the derived
structures must themselves be derived from the underlying physical
axioms. It follows that one cannot assign probabilities to the macro-world
branches as a physical axiom or make a primitive postulate specifying
that \textquoteleft we\textquoteright{} are necessarily in certain
branches; any such claim about the probabilities must be \emph{derived}
from the basic physical axioms, following either the deductive or
the inductive approach.

\subsection{The Case Against the Deductive Approach\label{subsec:The-Case-Against-the-Deductive-Approach}}

Among the existing formulations of the Many Worlds interpretation,
probabilities are typically not posited among their basic physical
axioms, in part because, as explained above, there is nothing fundamental
among the axiomatic physical ingredients of the Many Worlds interpretation
to which those probabilities could attach.

Moreover, as emphasized, for example, by Norton (2021)\nocite{Norton:2021tmtoi},
it is impossible to arrive deductively at a conclusion involving a
notion of probability without a notion of probability being inherent
in one\textquoteright s premises. Hence, \emph{a fortiori}, it is
not possible to arrive at the Born rule \emph{qua} probabilistic law
as the conclusion to a deductive argument solely from the axioms of
the Many Worlds interpretation alone.

A conceivable response to this objection would be to attempt to augment
the physical axioms of the Many Worlds interpretation with various
new physical axioms or with \textquoteleft reasonable\textquoteright{}
nonphysical principles \textendash{} such as some principles of decision
theory (Deutsch 1999, Wallace 2012)\nocite{Deutsch:1999qtopad,Wallace:2012temqtattei},
self-locating uncertainty (Vaidman 1998; Sebens, Carroll 2014)\nocite{Vaidman:1998oseotnowwsbitmwioqt,CarrollSebens:2014mwtbraslu},
or appeals to Boltzmannian statistical mechanics (Saunders 2021)\nocite{Saunders:2021bciteioqm}.
However, we will argue in this paper that no new physical axioms that
could achieve this would be consistent with the Many Worlds interpretation\textquoteright s
commitments, and that any non-physical principles one might seek to
use here would either be inconsistent with the physical axioms of
the Many Worlds interpretation and its vastly non-uniform ontology,
or would otherwise be impossible to justify.

\subsection{The Case Against the Inductive Approach\label{subsec:The-Case-Against-the-Inductive-Approach}}

The difficulty with the inductive approach is that any reasonable
invocation of induction, either in its logically implicative sense
or in its epistemically inferential sense (such as for abduction or
inference to the best explanation), requires an appeal to some sort
of \textquoteleft uniformity principle\textquoteright{} about nature.
But the picture of reality entailed by the Many Worlds interpretation,
if taken seriously, confounds any such uniformity principle. To be
more precise, if one takes the ontology of the Many Worlds interpretation
to its logical conclusions, then one sees that for almost any conceivable
kind of pattern of observable phenomena, there is, somewhere among
the infinitely many macro-world branches, a branch that realizes that
pattern. 

To make the inductive approach work out, one would therefore need
to postulate an additional, \emph{ad hoc}, \emph{sui generis} physical
axiom stipulating that \textquoteleft we\textquoteright{} \emph{qua}
physical beings belong to a macro-world branch with the right kind
of uniformity \textendash{} which is essentially equivalent to assuming
the Born rule outright. Such a physical axiom would need to refer
to non-fundamental physical ingredients of the Many Worlds interpretation
\textendash{} the specific physical observers that constitute \textquoteleft us\textquoteright{}
\textendash{} and thus such an axiom would necessarily run into the
same problems as we have already seen for the \textquoteleft axiomatic\textquoteright{}
or \textquoteleft deductive\textquoteright{} approaches, so such a
posit cannot be justified.

\section{The Axiomatic Approach\label{sec:The-Axiomatic-Approach}}

It has sometimes been proposed that the Born rule should simply be
included as an additional physical axiom of the Many Worlds interpretation,
sometimes known as the \textquoteleft Vaidman Rule\textquoteright{}
(Ridley, 2023)\nocite{Ridley:2023qpfts}, but it should be kept in
mind that it would need to be a very unusual kind of physical axiom. 

On all the currently available formulations of the Many Worlds interpretation,
the physical axioms describe the content of reality entirely in terms
of a \textquoteleft universal wavefunction\textquoteright{} evolving
through time according to a \textquoteleft unitary\textquoteright{}
dynamical law \textendash{} essentially, the Schrödinger equation
for some Hamiltonian. On some axiomatizations of the Many Worlds interpretation,
one includes further structure, such as a preferred collection of
self-adjoint operators with designated physical meanings (Wallace
2012)\nocite{Wallace:2012temqtattei}, and potentially also a classical
background spacetime with a family of mathematical operators (specifically,
reduced density matrices) assigned to different regions of space (Wallace,
Timpson 2010; Wallace 2012)\nocite{WallaceTimpson:2010qmosissr,Wallace:2012temqtattei}.

However, on none of these axiomatizations of the Many Worlds interpretation
do the physical axioms say anything at all about probabilities. So
if we want to posit the Born rule as a primitive physical axiom, along
with some attendant notion of probability, we are faced with the following
question: what, specifically, should the probabilities of the Born
rule refer to? What, specifically, should they be probabilities \emph{of}?

The issue here is that in modern versions of the Many Worlds interpretation,
the macro-world branches are not fundamental. Nor are the measuring
devices or external observers. And there are good reasons for that. 

Historically, one reason was to get around the problem that an abstract
universal wave function \emph{qua} vector fails to have a \emph{unique}
decomposition into a sum of vector-components serving as a preferred
basis. The universal wave function therefore fails to have a unique
set of branches to which to assign a notion of probabilities. This
\textquoteleft preferred basis\textquoteright{} problem is apparently
obviated by letting decoherence emergently determine a choice of branches
(Wallace, 2012)\nocite{Wallace:2012temqtattei}. Those branches would
then themselves be emergent features of physical reality as well.

But there is also a more compelling reason for treating the branches
as emergent. As argued by Everett himself (Everett 1956, 1957)\nocite{Everett:1956ttotuwf,Everett:1957rsfqm},
as well as by Wallace (2023)\nocite{Wallace:2023tsibaorqminube},
a supposed advantage of the Many Worlds interpretation over other
approaches is that it does not require adding any additional mathematical
machinery to the unitary quantum formalism. It is then ostensibly
much easier to accommodate quantum field theory and other applications
beyond straightforward non-relativistic systems, since we do not have
to figure out how to extend our new mathematical machinery to those
domains.

Thus it is important to the case for the Many Worlds interpretation
that the macro-world branches can be shown to emerge naturally as
a consequence of decoherence, rather than having to be added as part
of some kind of fundamental physical postulate. If advocates of the
Many Worlds interpretation were forced to add branches as a form of
fundamental ontology, then the Many Worlds interpretation would seem
to be no better off than \textquoteleft single-world\textquoteright{}
approaches that add additional mathematical structure in order to
pick out one of those branches as the \emph{actual} world. And in
that case, it would be difficult to understand why one should opt
for the extravagant ontology of the Many Worlds interpretation over
a single-world approach.

Consequently, it seems that any plausible Many Worlds approach would
need to avoid making reference to branches in its fundamental physical
posits. Similar points could be made about fundamental posits referring
to measuring devices or observers. But then it is simply impossible
to add the Born rule as a primitive physical axiom, because at the
level of the theory\textquoteright s fundamental ontology there is
nothing to which its probabilities could attach. 

Taking all these lines of reasoning together, we conclude that the
axiomatic route to the Born rule is not available for the Many Worlds
interpretation.

\section{The Deductive Approach\label{sec:The-Deductive-Approach}}

The deductive approach should begin with premises consisting of the
standard physical axioms of the Many Worlds interpretation, perhaps
together with some additional physical axioms, and/or \emph{a priori}
normative or epistemic principles that are at least minimally consistent
with those physical axioms, and then proceed through valid logical
reasoning to arrive at the Born rule as a conclusion. 

Now, by themselves, the standard physical axioms of the Many Worlds
interpretation only include a universal wave function, unitary evolution
for that universal wave function, and, on later formulations (e.g.,
Wallace 2012; Wallace, Timpson 2010)\nocite{Wallace:2012temqtattei,WallaceTimpson:2010qmosissr},
some preferred collection of operators, perhaps along with some assumption
of a classical background spacetime with a notion of regions being
assigned reduced density matrices. As such, none of the standard physical
axioms involve aleatory or objective probabilities, and, as noted
in the previous section, they simply do not have the kinds of fundamentalia
(e.g., primitive or axiomatic branches, or measuring devices or observers)
to which objective probabilities could attach.

It is a well-known point that there is fundamentally no way to arrive
at conclusions about numerical probabilities via a deductive argument
from premises that lack probabilistic concepts. As Norton puts it
(see e.g., Ch. 10 of Norton, 2021)\nocite{Norton:2021tmtoi}, whenever
one is confronted with a purported argument of this kind, it is just
a game to go through the argument and look for the logical elisions
or the question-begging. Much like with proposals for perpetual motion
machines, given adequate time, effort, and patience, one always eventually
finds the error.

Consequently, an axiom dealing with either objective probabilities
or some kind of subjective credences will necessarily have to be added
to the standard physical axioms of the Many Worlds interpretation
if we hope to be able to derive the Born rule.

As a first try, one might attempt to introduce a new physical axiom
positing a fundamental notion of objective chance that is relevant
specifically in the Many Worlds context. The goal would then be to
start from the interpretation\textquoteright s newly augmented physical
axioms and then arrive deductively at the Born rule as the correct
formula for capturing objective chances about measurement outcomes.

However, such an approach would surely take us out of the realm of
what could reasonably be called the Many Worlds interpretation. A
central ethos of the interpretation is that the universal wave function
always involves deterministically and that there are no other dynamical
processes that could include objective chances, and no other pieces
of ontology that could be behaving in an objectively chancy manner.
The upshot is that there is no way to deduce the Born rule in terms
of objective chances from the Many Worlds picture, even with additional
physical axioms.\footnote{Chen\textquoteright s \textquoteleft Wentaculus\textquoteright{} proposal
(Chen 2024)\nocite{Chen:2024twdmrmtaot} generalizes the universal
wave function to a density matrix of rank greater than one, with the
goal of introducing objective chances as parts of this \textquoteleft universal
density matrix.\textquoteright{} However, a Many Worlds version of
this proposal would lack the structure to provide a unique decomposition
of such a universal density matrix as a convex sum of projectors.
As a consequence, there would be no way to assign an axiomatically
probabilistic meaning to the non-negative coefficients of any such
convex-sum decomposition. Even if one were to try to select one convex-sum
decomposition, its projectors would not interfere at all with each
other under unitary time evolution. Thus, each projector would represent
a distinct possible universal wave function, evolving on its own and
developing its own macro-world branches via decoherence. The non-negative
coefficients in the convex-sum decomposition would then have no connection
to the lower-level Born-rule probabilities needed for the individual
branches.}

The other option is to add some kind of \emph{a priori} normative
or epistemic principle to the set of physical axioms. As a first pass,
one might try asserting that there is simply some kind of primitive
fact about the beliefs that we should have in a Many Worlds universe
\textendash{} in particular, that we should, \emph{a priori}, set
our credences in accord with the Born rule. Because this is a primitive
posit, the claim would not be that we should have these particular
quantitative credences for some \emph{reason}. The claim would instead
be that we should simply have these credences because the posit says
so, without any further justification.

There are a number of problems with this approach. 

First of all, we generally hold beliefs for reasons, or for the purpose
of achieving certain goals, but if the Born rule is just a primitive
numerical constraint on credences, then neither of these can be the
case for the credences it recommends. If a proponent of this sort
of \emph{a priori} principle were to declare \textquotedblleft You
should set your beliefs in accordance with the Born rule,\textquotedblright{}
and a skeptic simply declined to do so, then it would be hard to see
what the proponent could possibly say to induce a change in the skeptic\textquoteright s
position.

Second, such a principle would seem like a strange candidate for an
\emph{a priori} normative or epistemic principle, because it depends
so sensitively on contingent facts about ontology and laws \textendash{}
a principle of the form \textquoteleft you should set your credences
to match the mod-squared amplitudes of the wave function\textquoteright{}
simply would not make any sense in the vast majority of possible worlds
(on the usual philosophical notion of possible worlds), since most
of them do not have a branching wave function with amplitudes. And
insofar as one believes that there are any \emph{a priori} normative
or epistemic principles at all, it seems natural to think that they
would be very general principles which would apply broadly across
many possible worlds, rather than making specific and ineliminable
reference to some feature of one very particular kind of ontology.
Thus a primitive \emph{a priori} principle which refers specifically
to mod-squared amplitudes seems not very credible.

Finally, there is a methodological objection. If it is a legitimate
scientific move simply to impose primitive constraints that make detailed
reference to specific elements of an individual theory on what we
should believe, then it is unclear how any scientific theory could
be judged against any other. Two scientists could lay down conflicting
posits about what people should believe and neither of those posits
would be subject to any kind of test or critique, so science as we
know it would no longer be viable.

Thus if one is committed to a deductive derivation of the Born rule
from the Many Worlds interpretation, it seems inevitable that this
should be done by appeal to some rather more general normative or
epistemic principle such that, in the context of an Everettian universe,
following this general principle would to lead one to set credences
to match the mod-squared amplitudes, and hence yield agreement with
the Born rule. 

One important class of such deductive approaches attempts to constrain
our credences by appealing to symmetries (Zurek 2005, 2014; Sebens,
Carroll 2014)\nocite{Zurek:2005pfebsrfe,Zurek:2014qdcratroqj,CarrollSebens:2014mwtbraslu}.
Positing that credences should respect certain physical symmetries
might initially look more scientifically respectable than simply insisting
that credences should obey the Born rule, both because this approach
is much more general and because appealing to symmetries is cosmetically
similar to the kinds of reasoning that often succeed in single-world
scenarios in physics. For example, in the philosophical discussion
over the principle of indifference (Eva 2019)\nocite{Eva:2019poi},
it has commonly been argued that the right way to partition an outcome
space in order to apply an indifference principle is dictated by the
physical symmetries of the situation (van Fraassen, 1989)\nocite{VanFraassen:1989las}.

However, it should be emphasized that in such single-world cases,
it is not just any set of symmetries that are relevant. As discussed
already in the case of choosing marbles from a jar in Section~\ref{sec:The-Many-Worlds-Interpretation},
what matters in determining the appropriate credences is the set of
symmetries \emph{of the process that selects the outcome that we are
trying to predict}. It can be shown that setting our credences in
accordance with \emph{those} symmetries will, with high probability,
result in making accurate predictions, so there is a good reason why
we should set our credences in that way if we want to achieve specific
goals. But without knowledge or assumptions about the nature of the
selection process, there is nothing to distinguish between symmetries
which are relevant and symmetries which are not, so symmetries cannot
provide any meaningful guidance for rational credences. 

As another example, suppose we are trying to predict where a dart
will land on a dart board. The dart board itself has a symmetry under
rotations, so one might initially be tempted to assign a probability
distribution that is rotationally symmetric. But of course, what really
matters here is not the symmetries of the \emph{board} but rather
the symmetries of the way the darts are \emph{thrown}: if the darts
are thrown in a way that favors the upper left quadrant, for example,
it would not make sense to assign a probability distribution that
is rotationally symmetric. One might have an intuition that in
the absence of any knowledge of the throwing process we should opt
for the rotationally symmetric distribution, but this intuition
ultimately stems from implicit assumptions about how people throw
darts \textendash{} i.e., the distribution chosen is not \emph{a priori}
but comes from the \emph{empirical} observation that most people throw
in a way that is roughly rotationally symmetric and that individual
biases tend to be evenly distributed over throwers in ways that will
cancel out over a large enough ensemble. So there is no good reason
for assigning rotationally symmetric credences, or indeed any particular
distribution of credences, if we truly have no possible way to judge
or anticipate the likely symmetries or physical features of the way
the darts are being thrown. 

By contrast, in the Many Worlds case, there is no process that selects
the outcome that we are trying to predict \textendash{} there is no
\textquoteleft Cartesian ego\textquoteright{} randomly selecting from
among the branches at each moment. In effect, there is a dartboard
(i.e. a space of possible outcomes) but there is no one throwing any
darts. This means that although there may be symmetries around, they
are not symmetries relevant to the prediction task that we are interested
in, since they cannot possibly be symmetries of the process that selects
the outcome. In the dartboard analogy, the dartboard itself still
has a rotational symmetry in the absence of anyone throwing darts
at it, but because that symmetry is not associated with any selection
process there is no reason at all that it should be relevant to any
kind of credence assignation. So despite the superficial respectability
of these symmetry-based approaches, they are really no better motivated
than simply adding a primitive normative principle saying that we
ought to obey the Born rule.

Another important class of deductive approaches involve decision-theoretic
derivations of the Born rule (Deutsch 1999, Wallace 2012)\nocite{Deutsch:1999qtopad,Wallace:2012temqtattei}
based on \emph{a priori} normative principles of \textquoteleft rationality\textquoteright{}
or decision theory. The general idea is to posit what it means to
be a \textquoteleft rational agent,\textquoteright{} perhaps together
with various \textquoteleft background principles of scientific reasoning,\textquoteright{}
and deductively arrive at the Born rule as the proper way for rational
agents in a Many Worlds universe to set their credences.

Certainly it is at least minimally consistent with the physical axioms
of the Many Worlds interpretation to include additional principles
about what it means to be a \textquoteleft rational agent.\textquoteright{}
But if you ever meet anyone on the street \textendash{} say, a politician,
a religious evangelist, or an itinerant philosopher \textendash{}
who tells you that you ought to be rational, and that, furthermore,
you ought to be rational in a specific way, then you would have every
right to ask that person why. \textquotedblleft Why be rational in
your sense of the word, and why be rational in the specific way that
you suggest?\textquotedblright{}

You would be well within your rights to expect the person to give
you some reason, such as that being rational in this specific way
will increase your probability of some sort of success in your life.
But the problem is that such an answer can never be given in the Many
Worlds scenario, because we cannot show that some principle of \textquoteleft rationality\textquoteright{}
will increase the probability of success if we have not yet derived
a notion of probability. Thus no such principle can ever be used to
give a reason why we should treat Born rule weights as probabilities
in the first place.

Indeed, in the vastly branching ontology of the Many Worlds interpretation,
there will always be plenty of confounding agents similar to you up
to now who will go on to succeed in their futures despite persistently
violating whatever prescriptions of rationality the advocate \textendash{}
or really any sensible advisor \textendash{} would suggest that you
uphold. Like their namesake from the television show \emph{Seinfeld},
these \textquoteleft Costanza agents\textquoteright{} constantly do
the opposite of whatever good sense tells them to do, and yet, seemingly
miraculously, they still consistently succeed in their lives.\footnote{The name comes from the character George Costanza, played by Jason
Alexander, in a \emph{Seinfeld} episode titled \textquotedblleft The
Opposite\textquotedblright{} (Season 5, Episode 22), which originally
aired on May 19, 1994.} These agents are not acting rationally in the advocate\textquoteright s
sense, so why should you? For all you know, you might be a Costanza
agent yourself! In particular, you cannot appeal to the claim that
the probability of being a Costanza agent is low as part of the reason
why we should treat Born rule weights as probabilities, because no
notion of probability has been arrived at yet. 

One popular move to evade this sort of circularity is to appeal to
Dutch-book coherence (Wallace 2012)\nocite{Wallace:2012temqtattei}.
The idea here is to argue that if you fail to assign probabilities
to the outcomes of a betting game according to the Born rule, then
you will be guaranteed with certainty to lose money, without any need
to invoke any axiomatic probabilistic notions.

The immediate trouble with this line of argument is that it simply
fails to take the abundantly branching ontology of the Many worlds
interpretation sufficiently seriously. In addition to \textquoteleft maverick\textquoteright{}
macro-world branches that lie near the extreme edge of the distribution
of outcomes of the betting game in question (DeWitt 1970)\nocite{DeWitt:1970qmr},
there are also \textquoteleft super-maverick\textquoteright{} macro-world
branches that fail to stay within the bounds of the game altogether.
In particular, there are super-maverick branches in which the opponent
in the bet spontaneously ends up not Dutch-booking you, and the existence
of such branches means that you cannot, in fact, be guaranteed to
lose money. In particular, for the reasons already discussed, we cannot
use the claim that the super-maverick branches can be ignored or neglected
because they have low probability as part of the reason why we should
treat Born rule weights as probabilities.\footnote{It cannot even be taken for granted that agents will be in a mental
state of self-locating uncertainty immediately after measurements,
because there will be super-maverick branches in which their brains
are spontaneously in states of \emph{confidence} about their branch.
To assert that such unjustifiably confident agents can be ignored
because they are unlikely would again be a form of logical circularity.}

However, one desperate move might, in principle, appear to be available.
One could try to argue that super-maverick branches are guaranteed
to have such small amplitude, at the level of their Hilbert-space
norm, that, as \textquotedblleft emergent structures,\textquotedblright{}
they are not \textquotedblleft robustly present\textquotedblright{}
and are a mere \textquotedblleft trick of the light\textquotedblright{}
(Wallace, 2012, p. 253)\nocite{Wallace:2012temqtattei}. Then one
could rely ostensibly, once more, on Dutch-book coherence and its
prescription that acting on any other probability calculus than the
one based on the Born rule would lead to inevitable betting losses.

This move, however, could only succeed if super-maverick branches
were all \emph{assuredly} below some threshold of amplitude, because
if they \emph{ever} occurred above that threshold, in any contingent
circumstances, then it would be impossible to ensure a Dutch-book
as a general, non-contingent feature of the Many Worlds interpretation.
And if Dutch-book scenarios only obtained sometimes, but not always,
then we would be back to needing to invoke some sort of prior notion
of probability to say that they are \textquoteleft highly likely\textquoteright{}
to obtain.

Furthermore, relying on a \textquotedblleft threshold\textquotedblright{}
of \textquotedblleft robust existence\textquotedblright{} for macro-world
branches in the first place would then lead immediately to follow-up
Sorites-style questions about what the level of that threshold is,
and what sets the threshold at that level and not some other level.
These are all questions reminiscent of old debates over the Heisenberg
cut in the Copenhagen interpretation that was supposed to divide microscopic-quantum
phenomena from macroscopic-classical phenomena (Bell, 1990)\nocite{Bell:1990am}.
In any event, answering these sorts of questions would require adding
more physical axioms or principles of some kind, leading us back to
the same challenges described elsewhere in this paper.

It is true that other physical theories make approximations of various
kinds when describing emergent features. But justifying the claim
that a small perturbation should have small effects on our observations
is not a free-floating principle: it needs to be justified or accounted
for on each physical theory in question. 

For example, in Newtonian mechanics, when we approximate a slightly
oblate planet as a perfect sphere, we have a planet with a transparent
physical interpretation either way. The fact of having a planet in
the first place does not hinge on whether we can make the approximation
or not. Moreover, on the physical axioms of Newtonian mechanics \textendash{}
that the world consists of mereological sums of particles in three-dimensional
physical space obeying Newton\textquoteright s laws of motion \textendash{}
we can say just what that spherical approximation implies in terms
of changes to the behavior of the system, and we can also predict
what the theory tells us will be the resulting changes to our observations.
That is, we can compare what the Newtonian theory says our observations
would be with the approximation against what the Newtonian theory
says our observations would be without the approximation. One can
handle this difference in the predictions of our observations by using
the standard tools of perturbation theory.

On the Many Worlds interpretation, if we already knew that the amplitudes
of macro-world branches represented probabilities, then dropping branches
with tiny amplitudes would be a matter of convenience, because we
would have probabilities either way and the approximation would simply
mean approximating one probability as another. As in the Newtonian
case, we would be able to say definitively not just what the approximation
would imply for the changes to the behavior of the system, but also
what the theory would say would be the changes to our observations,
using perturbation theory as a tool for handling this comparison.

However, in the case of trying to account for Born-rule probabilities
in the first place, we cannot characterize the meaning of the approximation
as probabilistic \emph{a priori}, nor does the theory tell us what
the observational meaning of small perturbations in the amplitudes
and related approximations are. We cannot help ourselves to the reasoning
described above because it is simply not clear, on the Many Worlds
interpretation, how to compare what the theory says our observations
would be under the approximation against what the theory says our
observations would be without the approximation, because our observations
in the latter case are undefined. Specifically, whereas the approximation
of dropping the super-maverick branches would permit a Dutch-book-style
inevitable-loss argument and a seemingly clear prediction of our observations
in that case, without the approximation we would be confronted with
the super-maverick branches, and then the super-maverick branches
would make it unclear whether there is any sensible way to say what
our observations should be.

\section{The Inductive Approach\label{sec:The-Inductive-Approach}}

We will now examine the possibility that the Born rule could be justified
directly or indirectly by means of induction, broadly construed to
include abduction. 

As a first response, it is not necessarily obvious that induction
should be considered separately from the deductive route. The usual
application of induction in scientific practice is that we first arrive,
abductively, at some kind of theory of the world, and then we deduce
predictions about the future from that theory. This account of induction
suggests that induction can be applied only if there is some intermediate
theory that we can motivate via abduction, in which case we then return
to the problem discussed in the previous section about whether we
can get the Born rule deductively out of such a theory.

Indeed, as illustrated most vividly in Goodman's \textquotedblleft new
riddle of induction\textquotedblright{} (Goodman 1955)\nocite{Goodman:1955ffaf},
there is no obvious way to know which regularities should be extrapolated
into the future if we do not have some kind of organizing structure
in mind. An organizing structure of this kind is exactly what a good
scientific theory gives us. Moreover, trying to do induction without
a mediating theory seems particularly puzzling in the Many Worlds
context, because advocates of the interpretation presumably want to
use inductive reasoning to arrive at quantum theory itself, on the
Many Worlds axiomatization, so it is then unclear what rationale there
would be for adding an extra layer of induction on top without proceeding
deductively through the theory proper.

However, an advocate of the Many Worlds interpretation might question
this assumption about the way in which induction is used, arguing
that it is permissible to arrive inductively at beliefs about what
we are going to see in the future without appealing to a mediating
theory. This view would allow the Many Worlds advocate to argue that
we should expect to see Born-rule regularities in the future purely
because we have seen them in the past, without any need to explain
how this projection into the future arises from the theoretical machinery
of the Many Worlds picture. 

What is often left out of this sort of approach is the obvious follow-up
question: Why should the past success of the Born rule \textendash{}
or the past success of some more general criterion of rationality
that is supposed to imply the Born rule \textendash{} imply that we
should employ this form of rationality in the future? This implicit
last step relies on some form of inductive projection about the future
based on past patterns of phenomena or past experiences. 

Of course, it is well known that there is no foolproof argument that
provides a guarantee that inductive projections of this kind are going
to work. But in the more familiar, single-world case, there is at
least some way that the universe could be such that induction will
reliably work. Specifically, we assume that there is some sort of
regularity or uniformity principle of the right kind in the world
\textendash{} that the world contains regularities extending both
into the past and the future that observers like us are able to notice
and systematize. If this assumption is true, certain kinds of induction
will succeed (again, see Norton 2021)\nocite{Norton:2021tmtoi}. If
it is not true, induction will not succeed. 

Thus, we may think of this uniformity principle as a kind of necessary
condition for doing science. We can\textquoteright t be sure that
it\textquoteright s right, but we are entitled to assume it in the
practice of science because otherwise the practice of science would
be impossible. Indeed, such a uniformity principle, applied to the
\emph{entire universe as a whole}, is of course taken for granted
within the Everett approach as much as for any other scientific theory,
since the Many-Worlds approach holds that the universe is described
by a universal wave function that is consistently undergoing unitary
time evolution and that will continue doing so into the future. 

However, the problem is that on the Many Worlds picture, this standard
uniformity assumption by itself is not adequate to make any projections
about what we will see in the future, because nearly every pattern
of phenomena occurs in some branch, and only very rare branches have
the right kind of internal uniformity that conforms to the Born rule
or indeed allows any kind of inductive reasoning. So on the Many Worlds
picture, in order for induction to make sense, we would need an \emph{additional}
uniformity principle, over and above the one that grounds the unitary
evolution of the universal wave function, specifying that individual
observers in the Everettian universe will typically or usually or
with high probability see regularities in the future that look similar
to the regularities they have seen in the past. 

But then the problem is that, if it is accepted that there is no way
to derive or justify the Born rule deductively from within the theory
proper, then this additional uniformity principle just \emph{could
not possibly be true}. At any given point in the branching universe,
the possible branches lying ahead of an observer exhibit all possible
future regularities, regardless of what regularities that observer
has seen up until this point; and the same arguments showing that
there is no axiomatic or deductive way in this context to justify
the claim that we should expect to find ourselves in branches exhibiting
Born-rule regularities also demonstrate that there is no axiomatic
or deductive way to justify the claim that we should expect to find
ourselves in branches exhibiting regularities in the future similar
to those we have seen in the past. So the sort of inductive projection
about the future that we would need in order to account for the Born
rule is simply not consistent with the ontology of the Many Worlds
interpretation, according to which the vast majority of macro-world
branches fail to abide by the kind of uniformity principle that could
ensure the success of that inductive projection. 

This is important because there is a significant difference in rationality
between making the choice to take on some assumption that could possibly
be true in the relevant context because it is needed for the practice
of science, versus taking on some assumption that \emph{could not
possibly be true} in the relevant context. In the latter case it is
surely more reasonable to say that if the assumption in question is
truly essential to the practice of science in the relevant context,
then the practice of science just isn't possible at all in that context.
Thus, if the Many Worlds interpretation requires for its internal
rationality that we must accept an assumption about the world that
could not possibly be true given the commitments of the interpretation,
then we cannot accept the Many Worlds interpretation as a legitimate
outcome of scientific inquiry.

The only conceivable way out of this trap would be to try to combine
an axiomatic approach with an inductive approach. Specifically, we
would need to introduce an \emph{ad hoc}, bespoke physical axiom that
\textquoteleft we\textquoteright{} \emph{qua} physical beings live
in a high-weight macro-world branch whose past and future have the
right kinds of uniformity in their patterns of observable phenomena
to make the Born rule the right kind of systematization of the patterns
of phenomena in our branch. This additional physical axiom would give
us the kind of \emph{secondary}, \emph{branch-specific} uniformity
principle, over and above the uniformity that has already been assumed
for the unitarily evolving universal wave function \emph{as a whole},
that is needed to arrive at the Born rule.

That is, one would have to stipulate, as a basic physical axiom, that
\emph{you personally} are or will be in some particular region of
the branching structure. For example, one might try to posit that
you are one of the observers in a branch that exhibits relative frequencies
close to those predicted by the Born rule. Note that it is not good
enough merely to say that up until this point you have belong to such
a branch, because in order for the Born rule to be of any use, we
must postulate, as part of this physical axiom, that you will \emph{continue}
to be on such a branch in the future. Alternatively, we must understand
\textquoteleft you\textquoteright  to refer to some kind of temporally
extended existent (Saunders and Wallace, 2008)\nocite{SaundersWallace:2008bau}
and maintain that \textquoteleft you,\textquoteright  this temporally
extended existent, are in fact an existent that, in both the past
and future, occupies a branch exhibiting relative frequencies close
to those predicted by the Born rule.

Unfortunately, this sort of bespoke physical axiom runs immediately
into insurmountable problems.

First of all, one might think that stipulating, as an axiomatic matter,
which observer you are personally is simply not the kind of thing
that a scientific theory should be doing. 

Second, such an axiom could never receive any empirical support. For
the Everett framework says that, if you are somewhere in the middle
of a series of observations, then every possible sequence of past
events is always followed by every possible sequence of future events:
all the future possibilities always exist, and nothing about the branching
or the mod-squared amplitudes at a given point in the sequence is
changed by the facts about which relative frequencies you have seen
in the past. Thus it would seem that in the Everett picture, what
you have seen in the past is completely irrelevant to the question
of what branch you will be on in the future, so there does not seem
to be any observation you could ever make that would give you any
reason to think that you are or will be in a branch exhibiting some
particular set of properties \emph{in the future}. So again it is
unclear how we could reasonably come to accept a scientific theory
in which a posit of this kind plays an essential role, since nothing
could serve as evidence for it.

A further problem relates to the fact that science is done by a large
community of observers, and it is that whole community that has arrived
at quantum theory, complete with the Born rule. According to the Many
Worlds picture, by contrast, each of us is on our own individual journey
through the branching structure, so it is unclear how the results
of experiments performed by other people could somehow give me a reason
to expect that \emph{I personally} am located in a specific place
in the branching structure. Thus, if the Born rule is understood in
an \emph{individually} self-locating sense, then there would appear
to be a significant mismatch between the content of the theory and
the means by which it is arrived at.

Moreover, there is also a methodological objection: if we are allowed
to stipulate axiomatically that we are going to see certain relative
frequencies for no reason, then it's not clear why we should bother
with scientific theories at all. In particular, if the stipulations
are made for no reason at all, then there is also no reason for our
stipulations about the past to be connected in any specific way to
what we stipulate about the future, and thus making observations will
tell us nothing at all about what we should stipulate for the future.
So this kind of postulate seems quite hard to square with any reasonable
understanding of scientific practice.

Finally, not only would such a bespoke physical axiom complicate the
pristine and parsimonious physical axioms of the Many Worlds interpretation,
which already bakes in a uniformity principle at the higher level
of the \emph{entire universe as a whole}, but it would place observers
like us back into the fundamental physical axioms of quantum mechanics,
thus making illegal axiomatic statements about ingredients that are
not posited in the physical axioms. In particular, the whole reason
why we need a solution to the measurement problem in the first place
was because we think it is implausible that the basic physical axioms
of a theory should contain reference to observers (Bell 1981)\nocite{Bell:1981qmfc},
and thus if the Many Worlds approach cannot do without an axiom that
refers directly to observers, then it is simply not a solution at
all. 

Given the difficulties associated with simply extending the inductive
approach to the Born rule, advocates of the Many Worlds interpretation
might perhaps prefer to adopt a more specific \emph{abductive} approach
based on inference to the best explanation. Indeed, it is clear that
the Many Worlds interpretation \emph{without} some way of obtaining
the Born rule will explain almost nothing, so it does initially seem
plausible that we could justify the Born rule in such a way. For example,
if the truth of the Many Worlds interpretation is presupposed, then
one might think that we can look at the specific regularities that
have actually been observed and argue that the best available explanation
for those regularities, conditional on the truth of the Many Worlds
interpretation, is that there is some kind of principle requiring
that we should see high outcomes associated with high branch-weights.

However, there are two key problems here. First, for the reasons that
we have already discussed at length, it is not clear that adding the
branch-weight rule as a brute postulate counts as any sort of \emph{explanation}
of the observed statistics \textendash{} it is really just a mysterious
and unjustifiable stipulation, and it certainly does not seem to offer
any kind of improved understanding. And of course it certainly cannot
be the best explanation available if in fact it is not an explanation
at all!

Second, it should be kept in mind that the task here is not to justify
the Born rule conditional on the truth of the Many Worlds interpretation,
but rather to justify the Many Worlds interpretation itself. So to
use abduction as a solution in this case, it would be necessary to
argue that the whole package of the Many Worlds interpretation plus
the Born rule, considered together as a brute postulate, is the best
possible explanation of the observed statistics. And regardless of
what one thinks of the alternatives, it seems dubious that this could
possibly be the best explanation because, as noted, it is doubtful
that the brute postulate counts as an explanation at all. Certainly,
if it is an explanation, it does not seem like a very good one!

Note that this objection holds even if one genuinely believes that
no other existing explanation is any good, because it is commonly
suggested that inference to the best explanation is applicable only
in cases where at least one explanation seems at least somewhat satisfactory
\textendash{} it should not be applied in the case where no available
explanation is judged adequate (Lipton 1993, Lipton 2004)\nocite{Lipton:1993vitbge,Lipton:2004ittbe}.
Thus, if it is really true that the Many Worlds interpretation plus
the Born rule, combined as a brute postulate, is really our best option,
then the right move is not to believe the Many Worlds interpretation
but rather to go back to the drawing board and find a better option
that does actually furnish a meaningful explanation of the observed
statistics.

\section{A Fourth Way?\label{sec:A-Fourth-Way?}}

From the discussion above, it seems doubtful that we can reasonably
arrive at the Born rule in the Many Worlds interpretation either by
postulating it as an axiom, deriving it deductively from the theory,
or justifying it directly by induction. 

At this point it seems that the only remaining possibility would be
for the advocate of the Many Worlds interpretation to postulate a
\textquoteleft fourth way\textquoteright{} beyond the three methods
discussed here. That is, either the advocate must posit some entirely
new kind of bespoke or \emph{ad hoc} methodology that falls outside
of the usual methods of science, or the advocate must retrospectively
reinterpret some existing methods of science to say that, if the world
is described by the Many Worlds interpretation, then all along we
have been doing something other than what we imagine ourselves to
be doing. 

For example, one line of attack suggests that an appropriate semantic
analysis reveals that in an Everettian context, the sentence ``I
am uncertain about which outcome I will see'' uttered before a
branching event comes out as true despite the fact that we know some
version of this observer will see all of the possible outcomes, since
the word \textquoteleft I\textquoteright{} should be analyzed as
referring to an entire continuant rather than merely to the person-stage
present before the measurement (Saunders and Wallace, 2008)\nocite{SaundersWallace:2008bau}.
Now this analysis, if successful, yields only that there is a meaningful
notion of uncertainty in an Everettian context and thus does not get
us anywhere near the conclusion that this uncertainty should be quantified
by the Born rule specifically. But perhaps one might hope that further
semantic reinterpretations might be adequate to close this gap.

Although it is probably not possible to show conclusively that no
such \textquoteleft fourth way\textquoteright{} is viable, there are
surely reasons for skepticism. 

On the one hand, positing a completely new methodology seems quite
implausible, because what is needed here is to explain how we could
justifiably have arrived at the Many Worlds interpretation by means
of the methods that scientists have actually used, not how it could
be arrived at by a completely new method.

On the other hand, if we attempt to adopt the reinterpretation strategy,
then we run into some very strong constraints, because any reinterpretation
would have to be done in such a way as to preserve the empirical confirmation
attached not only to quantum theory but also to other existing parts
of science, and, in particular, to all of the auxiliary theories that
we have used to arrive at quantum theory in the first place. For example,
it would not be feasible to reinterpret the existing methods in a
way that undermines the empirical status of classical mechanics and
classical electromagnetism in their appropriate regime, because we
had to use classical mechanics and electromagnetism to build the measuring
devices that were used in the experimental tests of quantum theory.
These sorts of constraints place significant limitations on the scope
of possible reinterpretations (Adlam 2025)\nocite{Adlam:2025ssfqmteotmp}.

In light of all these obstructions, we are quite sceptical about the
existence of a fourth way, even if we allow novel semantic analyses
of certain words for the Many Worlds context. Certainly, until such
a thing is actually exhibited, one cannot rest an important claim
about nature on such hopes alone.

\section{Conclusion\label{sec:Conclusion}}

We close by noting some interesting directions for future work. 

First, although we did not assume here that the Born rule is a
law of nature, its universality and reliability do seem to make it
a plausible candidate for that status. Indeed, the properties that
we have attributed to it \textendash{} that it is reliable, robust,
quantitatively precise, and a predictive generalization of patterns
of observation about highly generic kinds of microscopic systems \textendash{}
could plausibly be regarded as sufficient conditions for lawhood,
if not all necessary conditions. It might therefore be worthwhile
to investigate in greater detail the links between the Born rule and
various philosophical accounts of lawhood.

Second, we note that an argumentative strategy similar to the one
that we have mounted here seems likely to apply more broadly across
philosophy. In particular, similar problems about justifying a specific
distribution of probabilities arise in discussions over the problem
of priors, principles of indifference and Bertrand's paradox, and
objective Bayesianism. These sorts of problems also show up in the
context of the cosmological multiverse, and, more generally, in any
area of philosophy that relies on the use of self-locating probabilities,
ranging from anthropic arguments to the simulation hypothesis.

A similar axiomatic-deductive-inductive trichotomy may be applied
in all these other examples. First, it is not possible in these kinds
of cases to derive credences or probabilities directly from the structure
of the physical world, and any additional principles that we might
add end up collapsing back into the brute-postulate approach. Second,
it is not clear that there are any truly \emph{a priori} principles
that constrain credences directly in the absence of any specification
of an epistemic goal \textendash{} to be meaningful, such constraints
should come from specifying an epistemic goal and showing how a certain
credence distribution helps achieve that goal. And third, induction
will not work in these kinds of cases because there is no possible
way the world could be that would justify any expectation of a connection
between the past and the future. So just as we have concluded that
there is no viable way to derive the probabilities in the Many Worlds
interpretation, we maintain that there is likely no viable way to
derive rationally compelling probabilities in these kinds of examples
either, and it would be worthwhile to see how the arguments apply
in these cases as compared to the Many Worlds interpretation discussed
here.

\bibliographystyle{1_home_jacob_Documents_Work_My_Papers_2026-Agai___n_in_Bohmian_Mechanics_custom-abbrvalphaurl}
\bibliography{0_home_jacob_Documents_Work_My_Papers_Bibliography_Global-Bibliography}

\end{document}